# Direct measurements of the extraordinary optical momentum and transverse spin-dependent force using a nano-cantilever

M. Antognozzi[1,2], C. R. Bermingham[1], R. L. Harniman[1,3], S. Simpson[1,4], J. Senior[1], R. Hayward[1], H. Hoerber[1], M. R. Dennis[1], A. Y. Bekshaev[5], K. Y. Bliokh[6,7], and F. Nori[6,8]

[1]*H.H. Wills Physics Laboratory, University of Bristol, Bristol BS8 1TL, UK*
[2]*Centre for Nanoscience and Quantum Information, University of Bristol, Bristol, UK*
[3]*School of Chemistry, University of Bristol, Bristol BS8 1TL, UK*
[4]*Institute of Scientific Instruments of the ASCR, Brno, Czech Republic*
[5]*I.I. Mechnikov National University, Odessa, 65082 Ukraine*
[6]*Center for Emergent Matter Science, RIKEN, Wako-shi, Saitama 351-0198, Japan*
[7]*Nonlinear Physics Centre, RSPE, The Australian National University, Canberra, Australia*
[8]*Physics Department, University of Michigan, Ann Arbor, Michigan 48109-1040, USA*

**Known since Kepler's observation that a comet's tail is oriented away from the sun, radiation pressure stimulated remarkable discoveries in electromagnetism, quantum physics and relativity[1,2]. This phenomenon plays a crucial role in a variety of systems, from atomic[3–5] to astronomical[6] scales. The pressure of light is associated with the momentum of photons, and it is usually assumed that both the optical momentum and the radiation-pressure force are naturally aligned with the propagation of light, i.e., its wavevector. Here we report the direct observation of an extraordinary optical momentum and force directed perpendicular to the wavevector, and proportional to the optical spin (i.e., degree of circular polarization). Such optical force was recently predicted for evanescent waves[7] and other structured fields[8]. It can be associated with the enigmatic "spin-momentum" part of the Poynting vector, which was introduced by Belinfante in field theory 75 years ago[9–11]. We measure this unusual transverse momentum using a nano-cantilever capable of femto-Newton resolution, which is immersed in an evanescent optical field above the total-internal-reflecting glass surface. Furthermore, the transverse force we measure exhibits another polarization-dependent contribution determined by the imaginary part of the complex Poynting vector. By revealing new types of optical forces in structured fields, our experimental findings revisit fundamental momentum properties of light and bring a new twist to optomechanics.**

Since Euler's studies of classical sound waves, the wave momentum is naturally associated with the propagation direction of the wave, i.e., the normal to wavefronts or the *wavevector*. This idea was mathematically formulated by de Broglie for quantum matter waves: $\mathbf{p} = \hbar\mathbf{k}$. In both classical and quantum cases, the wave momentum can be measured via the pressure force on an absorbing or scattering detector. In agreement with this, Maxwell claimed in his celebrated electromagnetic theory that "*there is a pressure in the direction normal to the waves*"[1]. However, pioneering works by Poynting introduced the electromagnetic momentum density as a cross product of the electric and magnetic field vectors[2,12]: $\mathcal{P} \propto \mathbf{E} \times \mathbf{B}$. Unlike the straightforward de Broglie formula, the Poynting momentum is not obviously associated with the wavevector $\mathbf{k}$. It is indeed aligned with the wavevector in the simplest case of a homogeneous plane electromagnetic wave. However, in more complicated yet typical cases of *structured* optical fields[13,14] (e.g., interference, optical vortices, or near fields) the direction of $\mathcal{P}$ can differ from the wavevector directions[7,8].

Notably, the origin of this discrepancy between the Poynting momentum and wavevector lies within the framework of relativistic field theory (Supplementary Information). The conserved momentum of the electromagnetic field is associated with the translational symmetry of spacetime via Noether's theorem[10,15]. Applied to the electromagnetic field Lagrangian, this theorem produces the so-called *canonical* momentum density $\mathbf{P}^{can}$. In the quantum-field framework, the canonical momentum generates spatial translations of the field, in the same way as the de Broglie formula is associated with the operator $\hat{\mathbf{p}} = -i\hbar\nabla$ generating translations of a quantum wavefunction. Therefore, the canonical momentum density of monochromatic optical fields is naturally associated with the *local wavevector* $\mathbf{k}^{loc}$ of the wave electric field, which is determined by the phase gradient normal to the wavefront[7,8,13–15].

However, resolving fundamental difficulties with the canonical stress-energy tensor (which is non-symmetric and gauge-dependent), in 1940 Belinfante added a "virtual" contribution to get this to agree with the usual electromagnetic stress-energy tensor (symmetric and gauge-invariant)[9–11,15]. In monochromatic optical fields, assuming the Coulomb gauge, Belinfante's addition to the electromagnetic momentum is a solenoidal edge current $\mathbf{P}^{spin} = \frac{1}{2}\nabla \times \mathbf{S}$ produced by the intrinsic *spin* angular momentum density $\mathbf{S}$ (i.e., the oriented ellipticity of the polarization) of the field. Due to its solenoidal nature, this *spin momentum* does not transport energy, and is usually considered as unobservable per se. In contrast to $\mathbf{P}^{can}$, the Belinfante spin momentum is determined by the *circular polarization* and *inhomogeneity* of the field rather than by its wavevector[7–11].

Thus, the well-known Poynting vector represents a sum of qualitatively-different canonical and spin contributions: $\mathbf{P}^{can} + \mathbf{P}^{spin} = \mathcal{P}$. Moreover, it is the Belinfante spin momentum that is responsible for the difference between the local propagation and Poynting-vector directions in structured light.

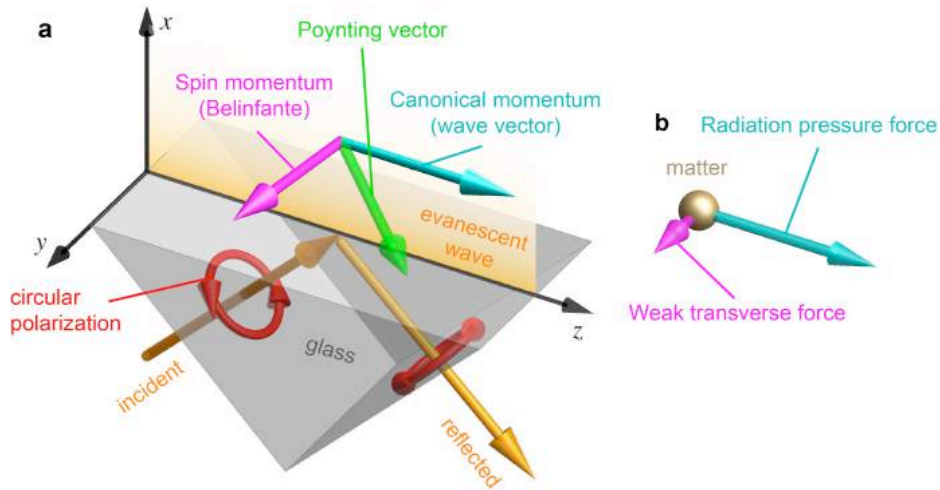

**Figure 1. Canonical and spin momenta of light in an evanescent wave. (a)** The evanescent wave is generated by the total internal reflection of a polarized plane wave at the glass-air interface. It carries longitudinal canonical momentum determined by its wavevector, and also exhibits transverse spin momentum, which is determined by the degree of circular polarization (helicity) of the field[7]. **(b)** The longitudinal canonical momentum produces the well-known radiation pressure in light-matter interactions, while the transverse spin momentum exerts a weak helicity-dependent force orthogonal to the propagation direction of light.

The above structure of the electromagnetic momentum has traditionally been regarded as an abstract field-theory construction. However, recently some of us argued[7] that one of the simplest inhomogeneous optical fields – a single evanescent wave – offers a unique opportunity to investigate, simultaneously and independently, the canonical and spin momenta of light in the laboratory environment, see Fig. 1. Considering the total internal reflection of a polarized plane wave at the glass-air interface, the canonical momentum density in the evanescent field in the air is proportional to its longitudinal wavevector: $\mathbf{P}^{can} \propto k_z \bar{\mathbf{z}}$. At the same time, the Poynting vector in an evanescent wave has an unusual *transverse* component, first noticed by Fedorov 60 years ago[16]. Remarkably, this component is proportional to the degree of circular polarization (helicity) $\sigma$ and has a pure Belinfante spin origin: $\mathcal{P}_\perp = \mathbf{P}_\perp^{spin} \propto \sigma \frac{\kappa k}{k_z} \bar{\mathbf{y}}$. Here $k$ is the vacuum wavenumber, $k_z > k$, and $\kappa = \sqrt{k_z^2 - k^2}$ is the parameter of the vertical exponential decay of the evanescent wave amplitude $\propto \exp(-\kappa x)$. Thus, if the spin momentum and Poynting vector are observable physical quantities, this should lead to an extraordinary helicity-dependent optical force, which is *orthogonal* to the propagation direction (wavevector) of the evanescent wave.

Here we present a direct measurement of the transverse helicity-dependent momentum and force in an evanescent wave, using a recently-developed atomic force microscope: the lateral molecular force microscope (LMFM)[17]. While conventional atomic force microscopes have the highest sensitivity to the vertical (i.e, normal to the interface) force component, the LMFM geometry, using a cantilever orthogonal to the surface, is ideal to measure the optical momenta parallel to the glass-air interface (see Fig. 2a). Similar sensors, perpendicular to a substrate, recently showed an extreme force resolution in various systems[17–20].

Importantly, the canonical and spin momenta of light manifest themselves very differently in light-matter interactions[7,8] (see Fig. 1b). The usual radiation pressure is produced by the *canonical* momentum (even though it is often attributed to the Poynting vector), and the corresponding force (also called the "scattering force") is always longitudinal, i.e., aligned with the wave propagation[7,8,14,21–23]: $\mathbf{F}_\parallel^{press} \propto \mathbf{P}^{can}$. In turn, the transverse spin momentum, in agreement with its "virtual" nature, can only produce a very *weak* force vanishing in the dipole-interaction approximation[7,8]: $\mathbf{F}_\perp^{spin} \propto \mathbf{P}_\perp^{spin}$, $|F^{spin}| \ll |F^{press}|$. In our experiment, we were able to significantly enhance the manifestation of the weak transverse force as LMFM uses a strongly *anisotropic* probe, which is highly sensitive to the optical force along one axis (Fig. 2). Namely, we used a planar dielectric nano-cantilever, which represents an ideal sensor for the force component normal to its plane[17–20]. Recently, there have been significant breakthroughs in the manufacturing of such highly compliant cantilevers, which are now truly nano-scale devices with femto-Newton sensitivity[19,20]. Mounting the cantilever in the $(x,z)$ plane of the evanescent wave (Fig. 1), one can measure the transverse $y$-component of the optical force.

We emphasize that the force we measure is neither the $z$-directed radiation-pressure (scattering) force[1–6,21–23], nor the gradient $x$-directed force used for optical trapping[3,4,21], but a novel type of optical force orthogonal to both the propagation and inhomogeneity directions. In contrast to the electric-dipole scattering and gradient forces, this weak force originates from the dipole-dipole coupling between electric and magnetic dipoles induced in matter, and in the generic case it contains two contributions proportional to the real and imaginary parts of the complex Poynting vector[7,8,24]. It is convenient to discriminate different types of optical forces via their dependence on the field polarization. Using the normalized Stokes-vector parameters $\vec{\mathcal{S}} = (\tau, \chi, \sigma)$, the radiation-pressure and gradient forces depend only on the first Stokes parameter $\tau$, while the weak transverse force has both the $\sigma$-dependent ($\mathbf{F}_\perp^{spin} \propto \mathbf{P}_\perp^{spin}$) and $\chi$-dependent ($\mathbf{F}_\perp^{Im}$, originating from the transverse "imaginary Poynting vector") contributions (Supplementary Information). In our experiment we observe both of these contributions, in agreement with recent theoretical predictions[7].

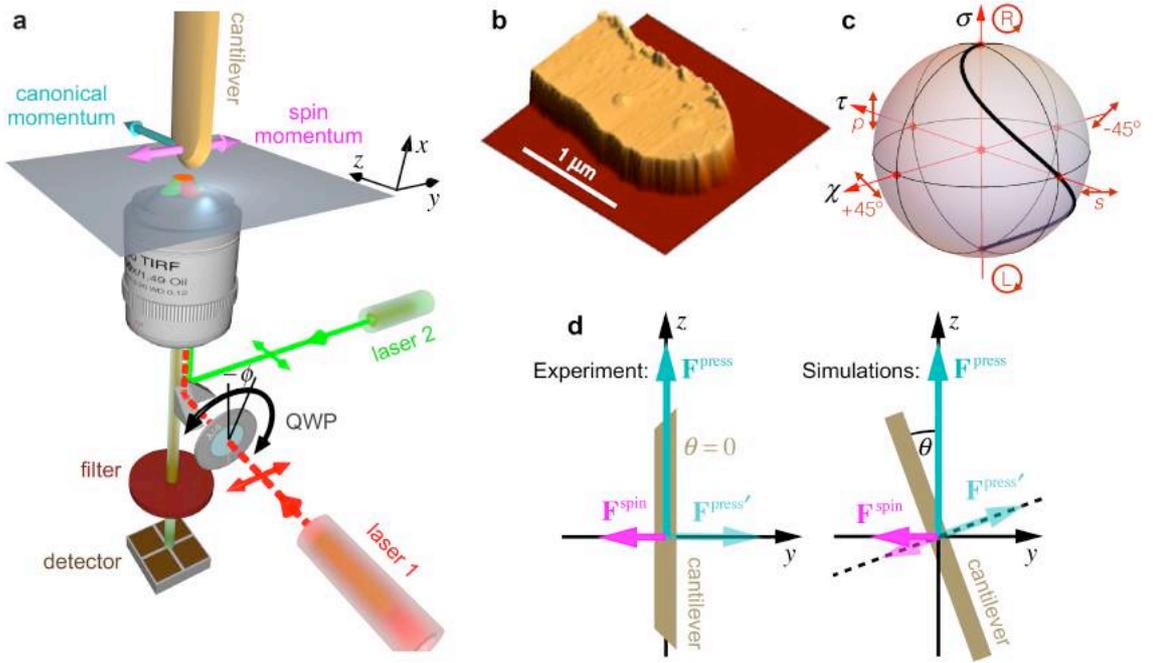

**Figure 2. Lateral molecular force microscope probing optical forces in an evanescent field. (a)** The LMFM setup. The red laser 1 produces the evanescent field to be probed. Its intensity is modulated and the polarization is controlled by a rotating quarter waveplate (QWP). The green laser 2 images the position of the cantilever probing the evanescent field of laser 1. **(b)** Atomic force microscope image of the free end of the cantilever. It has a complex shape with bevelled edges and surface inhomogeneities caused by the etching process. **(c)** Variations of the polarization state of the incident laser-1 field, caused by rotations of the QWP in the range of angles $-45° \leq \phi \leq 45°$, are represented by the black curve on the Poincaré sphere. The values $\phi = -45°$, $0°$, and $45°$ correspond to the right-hand circular (R), horizontal linear (*s*), and left-hand circular (L) polarizations, respectively. **(d)** Left: top view of the cantilever, whose shape has a $y \rightarrow -y$ asymmetry [cf. **(b)**]. This produces a transverse radiation-pressure force from the longitudinal canonical momentum of the field and mixes radiation-pressure and spin-momentum effects (cf., Fig. 1). Right: This mixing is modelled numerically using a symmetric cuboidal cantilever rotated by a small angle $\theta \ll 1$ about its vertical axis.

The experimental setup shown in Fig. 2a is based on the LMFM described in Ref. 20. The red laser 1 (wavelength $\lambda = 2\pi k^{-1} = 660$ nm) generates a $z$-propagating and $x$-decaying evanescent field at the glass-air interface via an objective-based total internal reflection system. The polarization state of this field is controlled by a quarter waveplate (QWP) with varying orientation angle $\phi$. Rotation of the QWP in the range of angles $-45° \leq \phi \leq 45°$ drives the polarization of the incident light between opposite spin states, i.e., between right-handed ($\sigma = 1$) and left-handed ($\sigma = -1$) circular polarizations on a path with nonzero $\tau$ and $\chi$ on the Poincaré $\vec{\mathcal{S}}$-sphere, as is shown in Fig. 2c. (Note that the polarization parameters of the evanescent wave differ slightly from those of the incident light, see Supplementary Information.) The cantilever, with a spring constant $\gamma \simeq 2.1 \cdot 10^{-5}$ N/m, is manufactured from ultra-low stress silicon nitride (refractive index $n = 2.3$); it has thickness $d \simeq 100$ nm, width $w \simeq 1000$ nm, and length $l \simeq 120$ μm (Fig. 2b). It is vertically mounted in the evanescent field, with its tip being 30 nm above the glass cover-slip. Deflections of the cantilever, $\Delta$, caused by optical forces, are registered using a detection system based on a non-interferometric scattered evanescent wave (SEW) method[20]. The SEW system involves the green laser 2 (wavelength 561 nm), and it allows the measurement, with a 1 nm resolution, of the cantilever deflections $\Delta$ as well as its vertical position. The intensity of the evanescent field produced by the red laser 1 is "on-off" modulated in time (TTL-modulation) to generate an intermittent force field. This allows us to isolate optical forces produced by the laser 1 on the constant background of other forces (e.g., from the imaging laser 2) (Fig. 3a).

An ideal cantilever with a symmetric cuboidal shape mounted in the $(x,z)$-plane would be insensitive to the longitudinal radiation pressure and would measure only the weak transverse force. However, the reactive-ion etching in the cantilever fabrication process results in an imperfect asymmetric shape with bevelled edges and varying surface roughnesses[19] (Fig. 2b). In particular, since the real cantilever has no mirror symmetry $y \to -y$, there is an asymmetric $y$-scattering of the $z$-incident light, producing a transverse scattering force which can be associated with the longitudinal canonical momentum of the field, Fig. 2d. Thus, the real cantilever measures the weak transverse force with an inevitable small admixture of the longitudinal radiation-pressure effect: $F^{\text{measured}} = F_\perp + \theta' F_\parallel^{\text{press}}$, where $\theta' \ll 1$ is an unknown parameter. However, these two contributions have different dependences on the wave polarization, which allows us to separate the different forces unambiguously. Indeed, the radiation-pressure (canonical momentum) force depends only on the first Stokes parameter $\tau$, and therefore is an *even* function of the QWP angle $\phi$. In turn, the weak transverse force has the $\sigma$-dependent (Belinfante spin momentum) and $\chi$-dependent ("imaginary Poynting vector") contributions, which are both *odd* functions of $\phi$ (Supplementary Information). Thus, the even and odd parts of the measured force $F^{\text{measured}}(\phi)$ correspond to the longitudinal radiation-pressure effects and the transverse weak force, respectively.

The results of our measurements are presented in Fig. 3. Figure 3a shows an example of the cantilever-position signal (detected via SEW by laser 2) varying in time due to the intermittent force produced by the laser-1 evanescent field. The distance $\Delta(\phi)$ between the centroids of the two Gaussian-like distributions, corresponding to the "on" and "off" laser 1, is a measure of the optical force: $F^{\text{measured}}(\phi) = \gamma \Delta(\phi)$. To improve the resolution and average out thermal fluctuations, we accumulated two distributions over 30 "on-off" cycles. The measured force $F^{\text{measured}}(\phi)$ versus the QWP angle $\phi$ is depicted in Fig. 3b. We neglect the $\phi$-independent contributions and plot the force with respect to its reference value at $\phi = 0$. It has a clearly asymmetric $\phi \to -\phi$ shape and different magnitudes for the right-hand and left-hand circular polarizations, which signals the presence of the $\phi$-odd spin-dependent transverse force. By retrieving the $\phi$-even and $\phi$-odd parts of $F^{\text{measured}}(\phi)$, we separate the radiation-pressure force (Fig. 3c) and the weak transverse force (Fig. 3d). The radiation-pressure force is proportional to the longitudinal canonical momentum dependent on the Stokes parameter $\tau$ ($F_\parallel^{\text{press}} \propto P_z^{\text{can}}$). In turn, analysing the $\phi$-dependence of the odd part, we find that it consists of both the $\sigma$-dependent transverse spin momentum ($F_\perp^{\text{spin}} \propto P_y^{\text{spin}}$) and $\chi$-dependent transverse "imaginary Poynting" ($F_\perp^{\text{Im}}$) contributions, as shown in Fig. 3d and predicted in theory[7]. These are the central results of this paper. They clearly show the presence of the transverse spin-dependent optical force, which is orthogonal to both the propagation and decay directions of the evanescent wave. This confirms the presence and observability of the enigmatic Belinfante spin momentum, which so far has been considered as "virtual". Furthermore, the measurements in Fig. 3b-d show that the spin momentum is indeed almost "invisible": the canonical-momentum contribution to the force is still five times stronger in our experiment despite its small weighting constant $\theta'$ (for an isotropic spherical particle it would be much stronger). These results prove that the Poynting vector, which has been used in optics for a century, does not present a single meaningful momentum of light, but rather a sum of two independent contributions of different nature and properties[7,8].

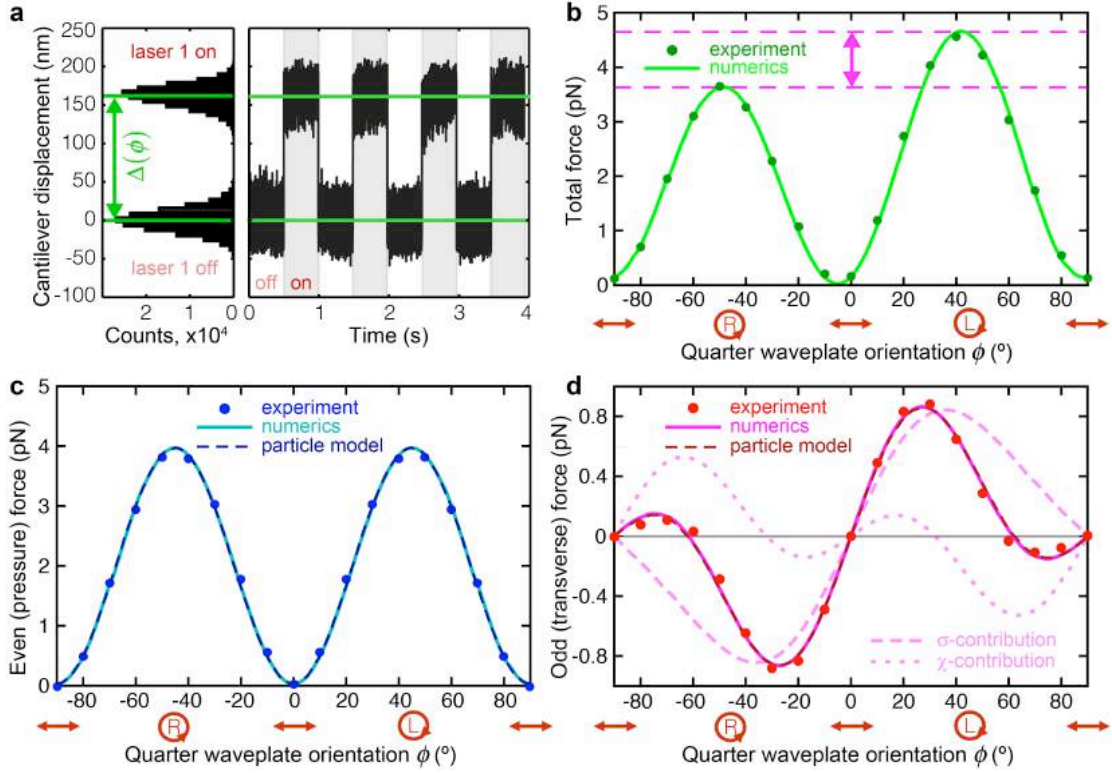

**Figure 3. Longitudinal and transverse optical forces in an evanescent wave. (a)** Right: typical cantilever-position trace, recorded at $\phi = -45°$, while the laser-1 intensity is TTL-modulated at 1 Hz. Left: the histogram of the position distribution shows two Gaussian-like distributions separated by a distance $\Delta(\phi)$. This yields the optical force acting on the cantilever: $F^{\text{measured}}(\phi) = \gamma \Delta(\phi)$. **(b)** The total force acting on the cantilever, $F^{\text{measured}}(\phi)$, as a function of the QWP angle $\phi$. **(c,d)** The longitudinal radiation-pressure force, $F_\parallel^{\text{press}}(\phi) - F_\parallel^{\text{press}}(0)$, and the weak transverse force $F_\perp(\phi)$, which are retrieved from the $\phi$-even and $\phi$-odd parts of the total force $F^{\text{measured}}(\phi)$. The transverse force includes the $\sigma$-dependent contribution $F_\perp^{\text{spin}} \propto P_y^{\text{spin}}$ from the Belinfante spin momentum, and also the $\chi$-dependent part $F_\perp^{\text{Im}}$ from the transverse "imaginary Poynting momentum". The experimental results are compared with the results of numerical simulations of the $\theta$-rotated cantilever (Fig. 2d) and calculations based on a simplified Mie-particle model (Supplementary Information). In panels **(b-d)**, the errors correspond approximately to the size of the symbols.

To verify our theoretical interpretation of the experimental measurements, we performed numerical simulations and analytical model calculations of optical forces on a matter probe in the evanescent field. Numerical simulations were performed using the coupled-dipole method, which models the cantilever as an assembly of interacting point particles (Supplementary Information). Since it is not practical to model the exact shape and inhomogeneities of the real cantilever, we used a simplified model of a cuboidal cantilever with the refractive index $n = 2.3$ and two geometric fitting parameters: (i) its thickness $d$, which controls the ratio of the $\sigma$- and $\chi$-contributions to the transverse force and (ii) a small orientation angle $\theta$, which controls the $y \to -y$ asymmetry of the cantilever and a small admixture of the $\tau$-dependent longitudinal radiation-pressure force (see Fig. 2d). The results of these simulations are shown as curves in Figs. 3b-d; they perfectly match the experimental data using only the common scaling factor and the fitting-parameters values $d \simeq 140$ nm and $\theta \simeq 0.08$ (i.e., $4.7°$). Moreover, the same $\tau$-dependent variations of the longitudinal force, as well as $\sigma$- and $\chi$-dependent transverse force, are obtained, using different scaling factors, within a greatly simplified model of a spherical Mie particle interacting with the field[7] (Supplementary Information). The main fitting parameter here is the particle radius, which is $r \simeq 139$ nm in our case. Importantly, the particle model provides *analytical* expressions for the forces, which confirm their direct proportionality to the canonical and spin momentum densities in optical fields[7,8]: $F_z^{\text{press}} \propto P_z^{\text{can}}$ and $F_y^{\text{spin}} \propto P_y^{\text{spin}}$ (Supplementary Information).

The numerical simulations also enabled us to investigate dependences of the radiation-pressure and transverse forces on the shape of the cantilever (see Supplementary Figure S5). In particular, varying the cantilever width $w$ (i.e., its area) we found that the longitudinal force $F_\parallel^{\text{press}}$ grows near-*linearly* with $w$, which reflects its usual *radiation-pressure* nature related to the planar surface of the cantilever. In contrast to this, the transverse force $F_\perp$ approximately *saturates* after $w$ reaches few wavelengths. This means that the weak spin-dependent force associated with the Belinfante spin momentum is not a pressure force, but rather an *edge effect* related to wave diffraction on the vertical edges of the cantilever. Indeed, one can show analytically that the transverse force *vanishes* for an infinite lamina without edges aligned with the $(x,z)$-plane: $F_y = 0$. This is in extreme contrast to the *infinite* radiation-pressure force for the same lamina in the $(y,z)$-plane: $F_z^{\text{press}} = \infty$. This proves that the spin momentum does not exert the usual radiation pressure on planar objects. Nonetheless, it can be detected (as we do in this work) due to its weak interaction with the edges of finite-size probes.

To conclude, our results re-examine one of the most basic properties of light: optical momentum and its manifestations in light-matter interactions. In contrast to numerous previous studies, which involved the radiation pressure in the direction of propagation of light or trapping forces along the intensity gradients, we have observed, orthogonal to both of these directions, the extraordinary optical momentum and force. Remarkably, the transverse Belinfante momentum and force are determined by the spin (circular polarization) of light rather than by its wavevector. Our results demonstrate that the canonical and spin momenta, forming the Poynting vector within field theory, manifest themselves very differently in interactions with matter. This offers a new paradigm for numerous studies and applications involving optical momentum and its manifestations in light-matter interactions[3–6].

Notably, the interplay between the canonical and Belinfante–Poynting momenta is closely related to fundamental quantum and field-theory problems, such as "quantum weak measurements of photon trajectories"[14,25], "local superluminal propagation of light"[14,22,23], and the "proton spin crisis" in quantum chromodynamics[26]. Furthermore, recently a reconstruction (but not direct measurement) of the *longitudinal* ($\sigma$-independent) Belinfante momentum was reported[27], which is associated with non-zero transverse spin density in structured fields[7,8]. In addition, there has been a strong interest in transverse spin-dependent optical forces near surfaces[28–30], which, however, originate from various particle-surface interactions rather than from pure field properties. All these studies reveal intriguing connections between (i) fundamental quantum-mechanical/field-theory problems involving optical momentum/spin and (ii) local light-matter interaction experiments with structured light fields. In this context, the LMFM technique used in our experiment offers a new platform for precision direction-resolved measurements of optical momenta and forces in structured light fields at subwavelength scales.

**Additional information.** Correspondence and requests for materials should be addressed to M.A. and K.Y.B.

**Acknowledgements.** This research was supported by Ministry of Education, Youth and Sports of the Czech Republick (project LO1212), RIKEN iTHES Project, MURI Center for Dynamic Magneto-Optics via the AFOSR (grant number FA9550-14-1-0040), Grant-in-Aid for Scientific Research (A), and the Australian Research Council. M.A. and R.L.H. would like to thank Nick and Susan Woollacott who kindly funded the equipment used in this research, as well as Adrian Crimp, Dave Engledew, Josh Hugo, and Pete Dunton for their essential technical support.

**Author contributions.** C.R.B., R.L.H. and S.S. contributed equally to this work. K.Y.B., M.R.D. and M.A. conceived the idea of this research. M.A. and R.L.H. designed the experiment. C.R.B. performed the measurements. S.S. performed numerical simulations. R.L.H. and C.R.B. contributed to the cantilever characterization. C.R.B., J.S. and R.H. collected and analysed data. H.H. contributed to the experimental protocol and methods for data analysis. A.Y.B. provided analytical and semi-analytical calculations of optical forces. K.Y.B. performed theoretical analysis and wrote the paper with input from M.A., S.S., M.R.D., A.Y.B. and F.N.

# SUPPLEMENTARY INFORMATION

# Direct measurements of the extraordinary optical momentum and transverse spin-dependent force using a nano-cantilever

M. Antognozzi[1,2], C. R. Bermingham[1], R. Harniman[1,3], S. Simpson[1,4], J. Senior[1], R. Hayward[1], H. Hoerber[1], M. R. Dennis[1], A. Y. Bekshaev[5], K. Y. Bliokh[6,7], and F. Nori[6,8]

[1]*H.H. Wills Physics Laboratory, University of Bristol, Bristol BS8 1TL, UK*
[2]*Centre for Nanoscience and Quantum Information, University of Bristol, Bristol, UK*
[3]*School of Chemistry, University of Bristol, Bristol BS8 1TL, UK*
[4]*Institute of Scientific Instruments of the ASCR, Brno, Czech Republic*
[5]*I.I. Mechnikov National University, Odessa, 65082 Ukraine*
[6]*Center for Emergent Matter Science, RIKEN, Wako-shi, Saitama 351-0198, Japan*
[7]*Nonlinear Physics Centre, RSPE, The Australian National University, Canberra, Australia*
[8]*Physics Department, University of Michigan, Ann Arbor, Michigan 48109-1040, USA*

## 1. Calculations of the fields and momentum densities

### *1.1. Canonical and spin momenta: from relativistic field theory to optical fields*

The canonical energy-momentum tensor for the free-space Maxwell field follows from the space-time translation symmetry of the field Lagrangian and Noether's theorem [10]. Using standard relativistic notation with the Einstein summation rule, the canonical energy-momentum tensor reads

$$T_{\text{can}}^{\alpha\beta} = \left(\partial^\alpha A_\gamma\right) F^{\beta\gamma} - \frac{1}{4} g^{\alpha\beta} F^{\gamma\delta} F_{\gamma\delta}, \tag{S1}$$

where $A^\alpha$ is the electromagnetic four-potential, $F^{\alpha\beta}$ is the field tensor, and $g^{\alpha\beta}$ is the Minkowski spacetime metric tensor. The tensor (S1) is gauge-dependent (due to the presence of $A^\alpha$) and non-symmetric. Nonetheless, it is this tensor that corresponds to the generators of spatial translations for the electromagnetic field.

In 1940, Belinfante suggested a symmetrisation procedure to "improve" tensor (S1), i.e., to make it gauge-invariant and symmetric [9,10]. He added the following total-divergence term (constructed from the spin tensor) to the canonical energy-momentum tensor:

$$T_{\text{spin}}^{\alpha\beta} = -\partial_\gamma \left( A^\alpha F^{\beta\gamma} \right), \tag{S2}$$

The resulting symmetric energy-momentum tensor (also known as the Belinfante energy-momentum tensor) is

$$\mathcal{T}^{\alpha\beta} = T_{\text{can}}^{\alpha\beta} + T_{\text{spin}}^{\alpha\beta} = F^\alpha_{\ \gamma} F^{\beta\gamma} - \frac{1}{4} g^{\alpha\beta} F^{\gamma\delta} F_{\gamma\delta}. \tag{S3}$$

This tensor (S3) is typically considered as meaningful in field theory, because it is gauge-invariant and is naturally coupled to gravity [10]. In turn, the Belinfante spin-correction term (S2) is usually regarded as "virtual", because it does not contribute to the energy-momentum conservation law, energy transport, and integral momentum of a localized field [10,11,15].

The momentum density of a free electromagnetic field is given by the $T^{i0} \equiv P^i$ components of the energy-momentum tensor. In this manner, the canonical, spin, and Poynting momentum densities are obtained from Eqs. (S1)–(S3) as

$$\mathbf{P}^{can} = \mathbf{E}\cdot(\nabla)\mathbf{A}, \quad \mathbf{P}^{spin} = -(\mathbf{E}\cdot\nabla)\mathbf{A}, \quad \mathcal{P} = \mathbf{P}^{can} + \mathbf{P}^{spin} = \mathbf{E}\times\mathbf{B}, \tag{S4}$$

where $\mathbf{E}$ and $\mathbf{B}$ are the electric and magnetic fields, whereas $\mathbf{A}$ is the vector-potential. Note that in Eqs. (S4) and in what follows we use Berry's notation $\mathbf{X}\cdot(\nabla)\mathbf{Y} \equiv X_i \nabla Y_i$ [13] and natural electrodynamical units $\varepsilon_0 = \mu_0 = c = 1$.

Although the canonical and spin momenta are gauge-dependent, there are several strong indications that in a number of situations the *experimentally-measured quantities correspond to the canonical quantities taken in some particular gauge*. Recently, this was actively discussed in relation to optical experiments with laser fields [7,8,14,15,22,23,25,S1–S6] and QED experiments detecting gluon and quark spin and orbital contributions to the proton spin [26]. In optical experiments, the measured quantities correspond to the *Coulomb gauge*, i.e., $A^0 = 0$ and $\nabla\cdot\mathbf{A} = 0$, and hereafter we assume this gauge.

We are interested in *monochromatic* optical fields, which can be described by complex time-independent field amplitudes: $\mathbf{E}(\mathbf{r},t) \to \operatorname{Re}[\mathbf{E}(\mathbf{r})e^{-i\omega t}]$, $\mathbf{B}(\mathbf{r},t) \to \operatorname{Re}[\mathbf{B}(\mathbf{r})e^{-i\omega t}]$, $\mathbf{A}(\mathbf{r},t) \to \operatorname{Re}[\mathbf{A}(\mathbf{r})e^{-i\omega t}]$, where $\omega$ is the frequency. Here we use the same letters $\mathbf{E}$, $\mathbf{B}$, and $\mathbf{A}$ for the complex field amplitudes, and imply only these complex fields in what follows. Importantly, the corresponding vector-potential amplitude (in the Coulomb gauge) becomes simply proportional to the electric field amplitude [15]: $\mathbf{A}(\mathbf{r}) = -i\omega^{-1}\mathbf{E}(\mathbf{r})$. Substituting these equations into Eqs. (S4), and performing *time averaging* over the $\omega$-oscillations, we obtain expressions for the canonical, spin, and Poynting momentum densities in a generic *optical* field [13,15]:

$$\mathbf{P}^{can} = \frac{1}{2\omega}\operatorname{Im}[\mathbf{E}^*\cdot(\nabla)\mathbf{E}], \quad \mathbf{P}^{spin} = \frac{1}{4\omega}\nabla\times\operatorname{Im}[\mathbf{E}^*\times\mathbf{E}], \quad \operatorname{Re}\mathcal{P} = \mathbf{P}^{can} + \mathbf{P}^{spin} = \frac{1}{2}\operatorname{Re}(\mathbf{E}^*\times\mathbf{B}). \tag{S5}$$

Here we use the same letters $\mathbf{P}^{can}$, $\mathbf{P}^{spin}$, and $\mathcal{P}$ for time-averaged optical momenta densities, and consider only these quantities in what follows. Note that for monochromatic fields the Poynting vector is described by a complex quantity $\mathcal{P} = (\mathbf{E}^*\times\mathbf{B})/2$, whose real part corresponds to the usual Poynting momentum (S5), whereas the imaginary part $\operatorname{Im}\mathcal{P}$ describes the "alternating flow of the stored energy" [12].

The canonical momentum in Eqs. (S5) is proportional to the local expectation value of the canonical momentum operator $\hat{\mathbf{p}} = -i\nabla$, and is proportional to the phase gradient or *local wavevector* of the field [13–15,23]:

$$\mathbf{P}^{can} \propto \operatorname{Re}[\mathbf{E}^*\cdot(\hat{\mathbf{p}})\mathbf{E}] \propto \mathbf{k}^{loc}|\mathbf{E}|^2. \tag{S6}$$

In turn, the spin momentum in Eqs. (S5) represents a solenoidal edge current, which is generated by the intrinsic spin angular momentum in the field:

$$\mathbf{P}^{spin} = \frac{1}{2}\nabla\times\mathbf{S}, \quad \mathbf{S} = \frac{1}{2\omega}\operatorname{Im}[\mathbf{E}^*\times\mathbf{E}] \propto \mathbf{E}^*\cdot(\hat{\mathbf{S}})\mathbf{E}, \tag{S7}$$

where $\hat{\mathbf{S}}$ is the vector of spin-1 matrices [13,15,S5]. Equations (S5)–(S7) reveal different physical origins and meanings of the two optical momenta constituting the Poynting vector.

Note that throughout this work we use the "standard" formalism for the canonical and spin quantities, which is based on the *electric* field [10,14,15,S3]. There is also a "dual-symmetric" formalism, where all quantities are symmetrized with respect to similar *electric* and *magnetic* field contributions [7,8,13,15,S3–S5,S7]. While the dual-symmetric approach is more natural for free-space fields, in our experiments the probes are sensitive only to the electric field, and, therefore, the standard "electric-biased" formalism is more suitable.

### *1.2. Fields and momenta in an evanescent optical wave*

To calculate the fields and momenta in the evanescent field investigated in our experiment, we need to describe several transformations of the optical field on its way from the laser to the probe (cantilever). The schematic of the experiment is shown in Figures 1 and 2 of the main text, and also in Figure S1.

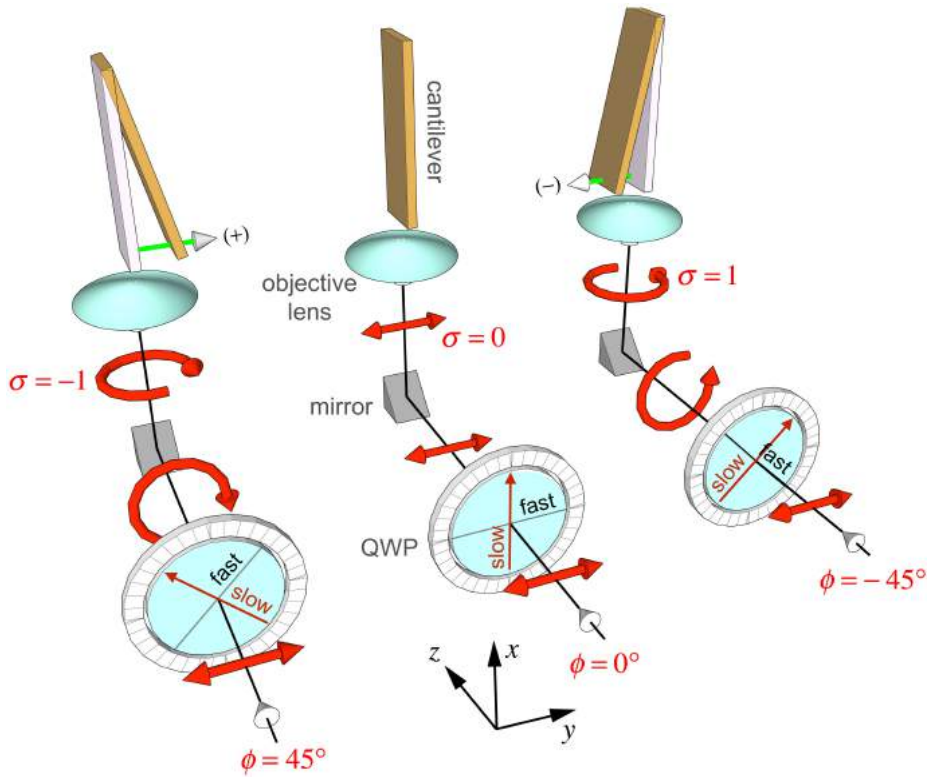

**Figure S1.** Transformations of the laser 1 beam and its polarization in the experimental setup (see also Fig. 2). Shown are: the coordinate frame $(x,y,z)$ accompanying the beam, rotations of the quarter waveplate (QWP) generating circular polarizations, flip of the circular polarizations after the beam reflection at the mirror, and spin-dependent deflections of the cantilever produced by the transverse force associated with the Belinfante spin momentum.

We describe the complex electric field of the laser-1 beam in its accompanying frame $(x,y,z)$, where the $z$-axis is directed along the beam and the $y$-axis is the transverse (horizontal) axis. Thus, the $s$ and $p$ polarizations correspond to the $y$ and $x$ linear polarizations, respectively. Neglecting the longitudinal $z$-component in the paraxial laser field, the laser 1 emits the $s$-polarized field $\mathbf{E}_0$ described by the following Jones-vector form:

$$\begin{pmatrix} E_{0x} \\ E_{0y} \end{pmatrix} = E_0 \begin{pmatrix} 0 \\ 1 \end{pmatrix} \exp(ikz). \tag{S8}$$

Here $E_0$ is the electric-field amplitude, and $k = 2\pi/\lambda$ is the wavenumber of the laser 1.

Next, the field (S8) is transmitted through the quarter waveplate (QWP) with its fast axis forming an angle $\phi$ with the $y$-axis, as shown in Fig. S1. (We intentionally defined the $\phi$ angle in the direction opposite to the usual rotations in the $(x,y)$-plane to account for the flip of the wave helicity in the reflection from the mirror after the QWP.) After the QWP, the slow-axis-polarized light acquires a $\pi/2$ phase shift with respect to the orthogonally-polarized component. As a result, the beam field $\mathbf{E}_1$ after the QWP is described by the following Jones-matrix transformation [S8]:

$$\begin{pmatrix} E_{1x} \\ E_{1y} \end{pmatrix} = \begin{pmatrix} \sin^2\phi + i\cos^2\phi & \sin\phi\cos\phi(1-i) \\ \sin\phi\cos\phi(1-i) & i\sin^2\phi + \cos^2\phi \end{pmatrix} \begin{pmatrix} E_{0x} \\ E_{0y} \end{pmatrix}. \tag{S9}$$

Substituting here Eq. (S8) and omitting the common $(1-i)/\sqrt{2}$ factor, we obtain

$$\begin{pmatrix} E_{1x} \\ E_{1y} \end{pmatrix} = \frac{E_0}{\sqrt{2}} \begin{pmatrix} \sin 2\phi \\ i + \cos 2\phi \end{pmatrix} \exp(ikz). \tag{S10}$$

After the QWP, the beam is redirected by the mirror, which works as an ideal reflector: the sign of the $y$-component of the field flips (which is equivalent, up to a total phase factor, to the $\phi \to -\phi$ transformation). Therefore, the field of the beam entering the objective lens becomes

$$\begin{pmatrix} E_{1x} \\ E_{1y} \end{pmatrix} = \frac{E_0}{\sqrt{2}} \begin{pmatrix} \sin 2\phi \\ -i - \cos 2\phi \end{pmatrix} \exp(ikz). \tag{S11}$$

Next, the beam enters the glass (a high-NA objective in our case, see Figs. S1 and S6 below) and undergoes a total internal reflection at the glass-air interface. The generation of the evanescent field and its properties at such interface are described in detail in the Supplementary Materials of [7]. Following that work, we represent the field (S11) inside the glass in three-dimensional form:

$$\mathbf{E}_1 \equiv \begin{pmatrix} E_{1x} \\ E_{1y} \\ E_{1z} \end{pmatrix} = \frac{E_0}{\sqrt{1+|m_1|^2}} \begin{pmatrix} 1 \\ m_1 \\ 0 \end{pmatrix} \exp(in_1 kz), \tag{S12}$$

where

$$m_1 = \frac{E_{1y}}{E_{1x}} = -\frac{i + \cos 2\phi}{\sin 2\phi} \tag{S13}$$

is the complex polarization parameter, and $n_1 = 1.5$ is the refractive index of the glass.

The plane-wave field (S12) impinges the glass-air interface with the angle of incidence $\alpha$, $n_1 \sin\alpha > 1$ ($\alpha \simeq 54.6°$ in our experiment). The evanescent wave field $\mathbf{E}$ generated in the air can be written as (see Supplementary Materials of [7])

$$\mathbf{E} \equiv \begin{pmatrix} E_x \\ E_y \\ E_z \end{pmatrix} = \frac{E}{\sqrt{1+|m|^2}} \begin{pmatrix} 1 \\ m k/k_z \\ -i\kappa/k_z \end{pmatrix} \exp(ik_z z - \kappa x), \qquad (S14)$$

where

$$k_z = k n_1 \sin\alpha \equiv k \cosh\vartheta, \quad \kappa = \sqrt{k_z^2 - k^2} \equiv k \sinh\vartheta, \qquad (S15)$$

are the propagation and exponential-decay wavevector parameters of the evanescent wave, which are expressed via the hyperbolic angle $\vartheta$. (In our experiment the propagation wavelength $2\pi k_z^{-1} \simeq 532$ nm and decay scale of the evanescent wave $\kappa^{-1} \simeq 150$ nm.) In Eq. (S14) we used the $(x,y,z)$ coordinates shown in Figures 1 and 2, and introduced the following amplitude and polarization parameters:

$$E = \frac{k}{k_z} \sqrt{\frac{1+|m|^2}{1+|m_1|^2}} T_p E_0, \quad m = \frac{T_s}{T_p} m_1, \qquad (S16)$$

which involve the Fresnel transmission coefficients [12]

$$T_s = \frac{2 n_1 \cos\alpha}{n_1 \cos\alpha + i \sinh\vartheta}, \quad T_p = \frac{2 n_1 \cos\alpha}{\cos\alpha + i n_1 \sinh\vartheta}. \qquad (S17)$$

Equations (S13)–(S17) completely describe the evanescent wave electric field $\mathbf{E}(\mathbf{r})$ that we probe in the experiment. Substituting this field into the general equations (S5)–(S7), we obtain the canonical and spin momentum densities in the evanescent wave:

$$\mathbf{P}^{can} = \frac{W}{\omega} k_z \left(1 + \tau \frac{\kappa^2}{k_z^2}\right) \bar{\mathbf{z}} = \frac{W}{\omega} k \left(\cosh\vartheta + \tau \frac{\sinh^2\vartheta}{\cosh\vartheta}\right) \bar{\mathbf{z}}, \qquad (S18)$$

$$\mathbf{P}^{spin} = \frac{W}{\omega} \left(-\frac{\kappa^2}{k_z} \bar{\mathbf{z}} + \sigma \frac{\kappa k}{k_z} \bar{\mathbf{y}}\right) = \frac{W}{\omega} k \left(-\frac{\sinh^2\vartheta}{\cosh\vartheta} \bar{\mathbf{z}} + \sigma \frac{\sinh\vartheta}{\cosh\vartheta} \bar{\mathbf{y}}\right). \qquad (S19)$$

We also determine the imaginary Poynting vector:

$$\mathrm{Im}\,\mathcal{P} = \frac{W}{\omega} \left(-\tau \frac{\kappa k^2}{k_z^2} \bar{\mathbf{x}} - \chi \frac{\kappa k}{k_z} \bar{\mathbf{y}}\right) = \frac{W}{\omega} k \left(-\tau \frac{\sinh\vartheta}{\cosh^2\vartheta} \bar{\mathbf{x}} - \chi \frac{\sinh\vartheta}{\cosh\vartheta} \bar{\mathbf{y}}\right), \qquad (S20)$$

which will play role in calculations of weak transverse optical forces. In equations (S18)–(S20) $W = \frac{1}{4}\left(|\mathbf{E}|^2 + |\mathbf{B}|^2\right) = \frac{1}{2} |E|^2 \exp(-2\kappa x)$ is the energy density of the field, $\bar{\mathbf{x}}$, $\bar{\mathbf{y}}$ and $\bar{\mathbf{z}}$ are the unit vectors of the corresponding axes, whereas

$$\tau = \frac{1-|m|^2}{1+|m|^2}, \quad \chi = \frac{2\,\mathrm{Re}\,m}{1+|m|^2}, \quad \sigma = \frac{2\,\mathrm{Im}\,m}{1+|m|^2} \qquad (S21)$$

are the Stokes parameters of the evanescent field [7]. In particular, the third Stokes parameter $\sigma$ is the degree of circular polarization, i.e., *helicity*, which determines the longitudinal $z$-directed *spin angular momentum* of light.

Substituting Eqs. (S13)–(S17) into Eq. (S21), we express the Stokes parameters of the evanescent wave as functions of the varying QWP angle $\phi$ and other parameters:

$$\tau = \frac{\sin^2 2\phi - (\cos^2 2\phi + 1)(\cosh^2 \vartheta - \cos^2 \alpha)}{\sin^2 2\phi + (\cos^2 2\phi + 1)(\cosh^2 \vartheta - \cos^2 \alpha)},$$

$$\chi = 2\frac{\sin 2\phi(-\cos 2\phi \cosh \vartheta \sin \alpha + \sinh \vartheta \cos \alpha)}{\sin^2 2\phi + (\cos^2 2\phi + 1)(\cosh^2 \vartheta - \cos^2 \alpha)},$$

$$\sigma = -2\frac{\sin 2\phi(\cosh \vartheta \sin \alpha + \sinh \vartheta \cos \alpha \cos 2\phi)}{\sin^2 2\phi + (\cos^2 2\phi + 1)(\cosh^2 \vartheta - \cos^2 \alpha)}. \tag{S22}$$

Importantly, the parameters $\tau(\phi)$, $\chi(\phi)$, and $\sigma(\phi)$ show different behaviour with variations of the QWP angle $\phi$, which enables us to separate the $\tau$-, $\chi$- and $\sigma$-dependent effects in our $\phi$-dependent measurements. In particular, $\tau(\phi)$ is an *even* function of $\phi$, while $\chi(\phi)$ and $\sigma(\phi)$ are *odd* functions. This yields a remarkable result: all *longitudinal* ($z$-directed) and *vertical* ($x$-directed) field properties in Eqs. (S18)–(S20) are *even* functions of $\phi$, while all *transverse* ($y$-directed) characteristics are *odd*. This allows an efficient separation of the usual in-plane and extraordinary transverse phenomena in $\phi$-dependent measurements. Note that the energy density $W(\phi)$ is also an even function of $\phi$:

$$W \propto |E_0|^2 \frac{2\cot^2 \alpha}{(n_1^2 - 1)} \left( \frac{\sin^2 2\phi}{\cosh^2 \vartheta - \cos^2 \alpha} + \cos^2 2\phi + 1 \right), \tag{S23}$$

which does not affect the above features of the polarization dependences.

It should be noticed that the polarization Stokes parameters in Eqs. (S18)–(S22) are those of the *evanescent* wave, and are determined via its polarization parameter $m$. This polarization and Stokes parameters are slightly different from those of the *incident beam*; the latters are described by Eqs. (S21) with the polarization parameter $m_1$. We use both kinds of Stokes parameters in this paper, to conveniently characterize the polarization at different stages. For instance, the polarization of the *incident* wave is depicted on the Poincaré sphere in Fig. 2c and indicated schematically under the $\phi$-axes in Figs. 3 and S2–S4. At the same time, discussing the $\tau$-, $\chi$-, and $\sigma$-contributions to optical momenta and forces, we imply the Stokes parameters of the evanescent wave.

Equations (S18)–(S20) show the presence of different kinds of optical momenta in the evanescent wave. The first one is the longitudinal $\tau$-dependent canonical momentum density $P_z^{\text{can}}$, proportional to the wavevector component $k_z$ [shown in the cyan frames in (S18)]. As $k_z > k = \omega/c$, this momentum density exceeds $\hbar\omega/c$ per photon. Therefore, the canonical momentum in an evanescent wave produces anomalously-high radiation pressure, which has been experimentally measured using light-atom interactions [22,14]. Second, the evanescent wave exhibits the spin momentum, which has a $\sigma$-*dependent component* $P_y^{\text{spin}}$, *perpendicular to both the propagation and decay directions of the wave* [shown in the magenta frames in (S19)]. This helicity-dependent transverse momentum is the main subject of our work. Finally, Eq. (S20) shows that the imaginary Poynting vector also has a transverse $y$-directed component $\text{Im}\mathcal{P}_y$ [shown in the light magenta frames], which is proportional to the second Stokes parameter $\chi$. As we show below (Section 2), this component also contributes to the weak transverse force on matter probes [7].

Note that the longitudinal component of the spin momentum in Eq. (S19), $P_z^{\text{spin}}$, produces only a weak $\tau$-dependent (i.e., $\phi$-even) correction to the strong radiation-pressure force from the canonical momentum, and, hence, can be ignored. Furthermore, its magnitude has a smallness of $\propto \kappa^2$ when $\kappa \ll k$. This longitudinal spin momentum is associated with another intriguing phenomenon: the transverse helicity-independent spin in evanescent waves [7,27,S5,S9–S13]. Similarly, the vertical component of the imaginary Poynting vector in Eq. (S20), $\mathcal{P}_x$, produces only a weak $\tau$-dependent correction to the strong vertical gradient force.

## 2. Calculations of optical forces and comparison with the experiment

Having the evanescent field (S13)–(S17) and its momentum properties (S18)–(S23), we now calculate how these properties reveal themselves in interactions with matter probes immersed in the evanescent wave. The momentum transfer in light-matter interactions produces optical forces, which we calculate below.

### 2.1. Calculations of the forces using the spherical-particle model

We first consider the simplest analytical model of a small spherical isotropic dielectric particle of radius $r \ll \lambda$ and permittivity $\varepsilon$ [7,8]. In the lowest orders in $kr \ll 1$, such a particle is characterized by the complex electric polarizability $\nu_e$:

$$\mathrm{Re}\,\nu_e = r^3 \frac{\varepsilon - 1}{\varepsilon + 2}, \quad \mathrm{Im}\,\nu_e = \frac{2}{3} k^3 \left(\mathrm{Re}\,\nu_e\right)^2, \tag{S24}$$

where the small imaginary part, $|\mathrm{Im}\,\nu_e| \ll |\mathrm{Re}\,\nu_e|$, originates from the radiation-friction effects [24,S14,S15]. The lowest-order magnetic polarizability of the particle is [S14]

$$\nu_m = k^2 r^5 \frac{\varepsilon - 1}{30}, \tag{S25}$$

so that $|\mathrm{Re}\,\nu_m| \ll |\mathrm{Re}\,\nu_e|$ and $\mathrm{Im}\,\nu_m \simeq 0$.

In the leading-order electric-dipole approximation, the optical force is given by [7,8,21,S4,S6,S15,S16]:

$$\mathbf{F} \propto \mathrm{Re}(\nu_e)\nabla W_e + \underbrace{\frac{1}{2}\omega\,\mathrm{Im}(\nu_e)\mathbf{P}^{\mathrm{can}}}_{\mathbf{F}^{\mathrm{press}}}. \tag{S26}$$

Here the first term is the *gradient* force $\mathbf{F}^{\mathrm{grad}}$ involving the electric energy density $W_e = |\mathbf{E}|^2 / 4$, and the second term is the *radiation pressure* force $\mathbf{F}^{\mathrm{press}}$ proportional to the *canonical momentum density* $\mathbf{P}^{\mathrm{can}}$. In the evanescent wave under consideration, the gradient force is directed along the vertical $x$-axis of the exponential decay, i.e., along the normal to the interface, while the radiation pressure is associated with the longitudinal $z$-direction of propagation and the canonical momentum density (S18).

Calculating the next-order correction to the electric-dipole force (S26), one can obtain a weak force, which originates from the dipole-dipole interaction between the electric and magnetic polarizabilities of the particle [7,8,24,S6]:

$$\delta\mathbf{F} \propto -\frac{\omega k^3}{3}\left[\mathrm{Re}(\nu_e \nu_m^*)\mathbf{P}^{\mathrm{can}} + \underbrace{\mathrm{Re}(\nu_e \nu_m^*)\mathbf{P}^{\mathrm{spin}}}_{\mathbf{F}^{\mathrm{spin}}} + \underbrace{\mathrm{Im}(\nu_e \nu_m^*)\mathrm{Im}\,\mathcal{P}}_{\mathbf{F}^{\mathrm{Im}}}\right]. \tag{S27}$$

The first term, proportional to $\mathbf{P}^{\mathrm{can}}$, is only a small correction to the radiation-pressure force in Eq. (S26), and it can be ignored in our considerations. The second term, proportional to $\mathbf{P}^{\mathrm{spin}}$, is the *weak spin force* $\mathbf{F}^{\mathrm{spin}}$ *associated with the Belinfante spin momentum*. It is mostly *transverse*, i.e., $y$-directed in the evanescent wave with $\kappa \ll k_z$. Note that for dielectric particles $\mathrm{Re}(\nu_e \nu_m^*) > 0$, and this force is *negative*, i.e., antiparallel to the spin-momentum direction. For non-absorbing dielectric particles, $k^3|\mathrm{Re}(\nu_e \nu_m^*)| \propto k^5 r^8$ and $|\mathrm{Im}\,\nu_e| \propto k^3 r^6$, so that $|\mathbf{F}^{\mathrm{spin}}|/|\mathbf{F}^{\mathrm{press}}| \propto (kr)^2 \ll 1$. Finally, the third term in (S27) is proportional to the *imaginary*

*Poynting vector* $\text{Im}\mathcal{P}$, and it also has the transverse $y$-directed component. For small particles with $kr \ll 1$, this force is much weaker than the spin-momentum force, $|\mathbf{F}^{\text{Im}}| \ll |\mathbf{F}^{\text{spin}}|$, because $k^3 |\text{Im}(v_e v_m^*)| \propto k^8 r^{11}$. However for larger objects with $kr \sim 1$ (which is the case in our experiment), the force $\mathbf{F}^{\text{Im}}$ also becomes noticeable.

In this work, we measure: (i) the longitudinal radiation-pressure force $F_z = F_z^{\text{press}} \propto P_z^{\text{can}}$ and (ii) the weak transverse force $F_y$, which includes the spin force $F_y^{\text{spin}} \propto P_y^{\text{spin}}$ and the imaginary-Poynting contribution $F_y^{\text{Im}} \propto \text{Im}\mathcal{P}_y$. Importantly, using the polarization dependences of these forces in the evanescent wave, Eqs. (S18)–(S23), we can clearly separate the above contributions even if they are mixed by a complex-shape probe (such as a cantilever). Namely, all the longitudinal (and also vertical) forces in Eqs. (S26) and (S27) depend only on the *first Stokes parameter* $\tau$, and therefore are *even* functions of the QWP angle $\phi$. In turn, the transverse spin force and transverse imaginary-Poynting contribution are proportional to the *third Stokes parameter* $\sigma$ (helicity) and the *second Stokes parameter* $\chi$, respectively. Therefore, both these parts of the transverse force are *odd* functions of the QWP angle $\phi$. Thus, the $\phi$-even and $\phi$-odd parts of the total force $F(\phi)$ measured by the cantilever, $F^{\text{even}}(\phi) \equiv [F(\phi) + F(-\phi)]/2$ and $F^{\text{odd}}(\phi) \equiv [F(\phi) - F(-\phi)]/2$, correspond to the longitudinal and transverse forces in the particle model.

In spite of its highly simplified character, the spherical-particle model can be used to characterize optical forces measured by the complex-shape cantilever in our experiment. Here we have to deal with not-small particles $kr \sim 1$ (because the cantilever thickness $d$ is not small: $kd \sim 1$). Therefore we use exact Mie-particle calculations [7], which take into account higher-order corrections to the electric and magnetic polarizabilities (S24) and (S25) but do not affect the general proportionality to the field momenta and the corresponding polarization parameters, Eqs. (S26) and (S27). The dielectric constant of the particle is set as that of the cantilever: $\varepsilon = n^2 = 5.3$. To compare the particle model and cantilever measurements, we involve three fitting parameters. First, the radius of the particle, $r$, is the main free parameter of the model. It controls the ratio between the $\sigma$-dependent and $\chi$-dependent contributions in the transverse force $F_y$. Second, since the weighting factors of the longitudinal and transverse forces mixed in the cantilever measurements depend on the shape effects, and also the total wave intensity is unknown, we introduce the scaling coefficients $K_{\parallel,\perp}$ between the calculated and measured quantities. As we are interested in the $\phi$-dependences of the forces, we consider the difference between the maximum and minimum values of the longitudinal radiation-pressure force $F_z^{\text{press}}(\phi)$:

$$\{\max[F^{\text{even}}(\phi)] - F(0)\}_{\text{measured}} = K_{\parallel} \{\max[F_z^{\text{press}}(\phi)] - F_z^{\text{press}}(0)\}_{\text{calculated}}, \tag{S28}$$

$$\{F^{\text{odd}}(\phi)\}_{\text{measured}} = K_{\perp} \{F_y(\phi)\}_{\text{calculated}}. \tag{S29}$$

The results of the comparison between the particle-model calculations and experimentally-measured forces are shown in Fig. S2. Using the scaling relation (S28), the calculated longitudinal radiation-pressure force (S26), $F_z^{\text{press}}(\phi) - F_z^{\text{press}}(0)$, is in very good agreement with the measured one, independently of the particle radius $r$. This is because the $\phi$-dependence of this force is determined by the first Stokes parameter $\tau(\phi)$, independently of $r$. At the same time, the $\phi$-dependence of the calculated transverse force (S27), $F_y(\phi)$, depends on the radius and shows the best agreement with the experiment for $r \simeq 139$ nm (i.e., comparable with the cantilever thickness $d \simeq 100$ nm). This corresponds to the value $kr \simeq 1.32$, for which both the spin contribution $F_y^{\text{spin}}$, proportional to $\sigma(\phi)$, and the imaginary-Poynting contribution $F_y^{\text{Im}}$, proportional to $\chi(\phi)$, play a role. The perfect agreement between the polarization $\phi$-dependences of the experimentally-measured and calculated forces proves that the measured even and odd parts of the total cantilever force can indeed be associated with the longitudinal and transverse optical forces. In turn, these forces are determined by the canonical momentum $P_z^{\text{can}}$ and transverse Belinfante momentum $P_y^{\text{spin}}$, together with the imaginary Poynting momentum $\text{Im}\mathcal{P}_y$, in the evanescent wave. Such robust polarization dependences of different forces, independently of the probe shape, confirm that we deal with intrinsic field properties. This is in contrast to "extrinsic" spin-dependent transverse-force effects, which originate from the probe-interface interactions [28–30] or specific properties of the probe, such as chirality [S4,S17–S19]. Note that in the case of chiral probes even the usual dipole radiation-pressure and gradient forces become spin- and helicity-dependent [S4,S17,S20–S22].

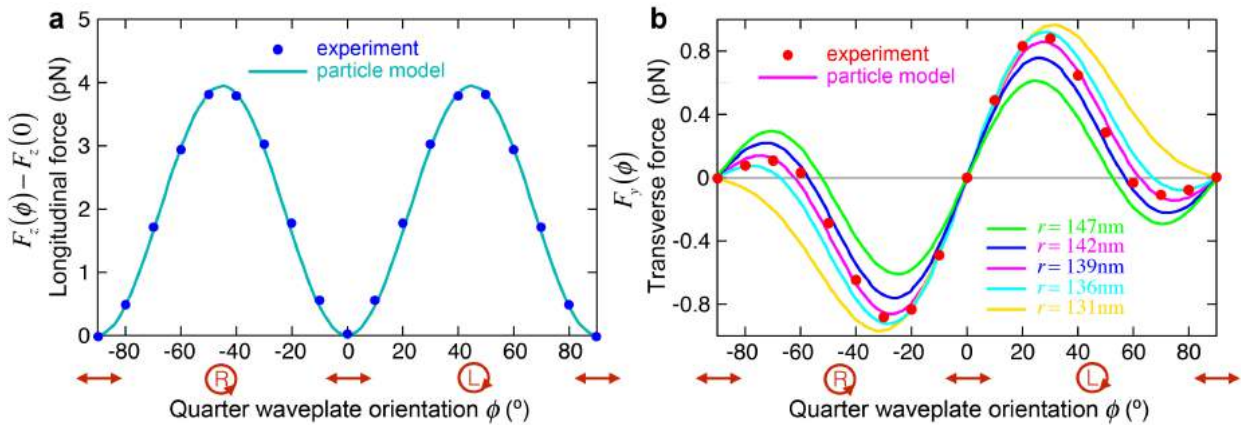

**Figure S2.** Longitudinal **(a)** and transverse **(b)** optical forces in an evanescent wave, calculated for a spherical particle, Eqs. (S26) and (S27), and fitted to the even and odd parts of the experimentally-measured force on the cantilever. The fitting procedure includes the scaling (S28) and (S29) and the fitting of the particle radius.

It is worth noticing that in the isotropic spherical-particle model, the longitudinal radiation-pressure force has the same order of magnitude for the *p*- and *s*-polarizations of light ($\tau = 1$ and $\tau = -1$). In contrast, for a cantilever with a highly-anisotropic vertical shape, the radiation-pressure force becomes very small for the horizontally *s*-polarized light. This difference between the particle and cantilever probes is not seen in our plots because we scale the variations of the radiation-pressure with respect to the *s*-polarized light ($\phi = 0$), Eq. (S28).

## 2.2. Numerical calculations of the forces for the cantilever

For numerical simulations of the interaction between the evanescent optical field and the cantilever, we employ the Coupled Dipole Method (CDM) [S23,S24]. In this method, continuous matter objects are decomposed into cubic arrays of point polarizable dipoles (cells) coupled through the electromagnetic dipole interaction tensor. When exposed to an external field, each cell feels not only the incident field but also the field scattered by all the other cells in the structure. Algebraically, this results in a large set of linear equations whose solution yields the polarization of each cell. Our implementation of this method for the simulation of the mechanical action of light is discussed elsewhere [S25], and we use the Quasi-Minimal Residual Method (QMR) [S26] with matrix-vector multiplication accelerated by fast Fourier transforms [S27].

The total (incident plus scattered) field outside the matter object is then given by the sum of the incident field with all of the radiating dipole fields. As the numerical lattice is refined, the total field converges to the continuum result. This approach lends itself well to structures of extreme geometry, such as the cantilever used in our experiment. Once the total field is known, the time-averaged optical force acting on each cell can be found by calculating the flux of the Maxwell stress tensor [S14,S24,S25,S28,S29].

In our simulations, the incident evanescent field is given by Eqs. (S13)–(S17) with the corresponding parameters of the experiment, while the cantilever is modelled as a dielectric cuboid with permittivity $\varepsilon = n^2 = 5.3$. If not otherwise stated, we use the following geometric parameters of the cantilever: thickness $d = 140$ nm (obtained from the fitting procedure described below), width $w = 1000$ nm, and infinite vertical length $l \to \infty$. Convergence of the model has been achieved by (i) refining the numerical lattice, (ii) increasing the vertical length $l$ of the cantilever. The resulting model uses $N \sim 2.5 \cdot 10^6$ cells and $l = 500$ nm (this value approximately equals to the triple decay length $3\kappa^{-1} \simeq 450$ nm, and further increase of $l$ practically does not affect the simulation results).

To characterize the $y \to -y$ asymmetry of the real cantilever (Figs. 2b,d), which couples the longitudinal radiation pressure to the transverse direction, we rotate the ideal symmetric (cuboidal) cantilever in numerical simulations by a small angle $\theta \simeq 0.08$ (i.e., 4.7°) with respect to the $z$-axis in the $(y,z)$-plane, as shown in Fig. S3a. We numerically calculate the force $\mathbf{F}^{\text{numeric}}(\phi) = F^{\text{numeric}}(\phi)\bar{\mathbf{n}}$ directed along the normal $\bar{\mathbf{n}}$ to the cantilever surface, which mixes the weak transverse force $F_y(\phi)$ and the radiation pressure $F_z^{\text{press}}(\phi)$ contributions. Using two fitting parameters: (i) the cantilever thickness $d$ and (ii) its orientation angle $\theta$, we fit the numerically-calculated force $F^{\text{numeric}}(\phi)$ to the measured force $F^{\text{measured}}(\phi)$ using only the common scaling factor (as the light intensity is unknown). With the values $\theta \simeq 0.08$ and $d = 140$ nm we achieve perfect agreement between the numerical simulations and experimental measurements, as shown in Fig. S3b.

Next, we decompose the total calculated force $F^{\text{numeric}}(\phi)$ into three contributions, proportional to the polarization Stokes parameters $\tau$, $\chi$, and $\sigma$ (the small polarization-independent contribution is ignored). First, the $\tau$-dependent contribution is associated with the radiation-pressure force $F^{\text{press}}$ (S26) proportional to the $\tau$-dependent canonical-momentum $P_z^{\text{can}}$ (S18). This force is an even function of the QWP angle $\phi$, and it is shown in Fig. S3c. Second, the $\chi$- and $\sigma$-dependent contributions constitute the weak transverse force (S27) [7]; their odd $\phi$-dependences are shown in Fig. S3d. The helicity-dependent $\sigma$-contribution is associated with the transverse Belinfante spin momentum in the evanescent wave: $F_y^{\text{spin}} \propto P_y^{\text{spin}}$ (Fig. 1). At the same time, the $\chi$-dependent part is the transverse force associated with the imaginary Poynting vector: $F_y^{\text{Im}} \propto \text{Im}\,\mathcal{P}_y$. It is also clearly present, both in simulations and measurements because the cantilever thickness is not small: $kd \sim 1$. The presence of the $\chi$-dependent contribution to the transverse force allows to discriminate between the "intrinsic" transverse force predicted in [7,8], which originates from the inherent field properties, and "extrinsic" spin-dependent transverse forces which arise from the coupling between the probe and an interface [28–30].

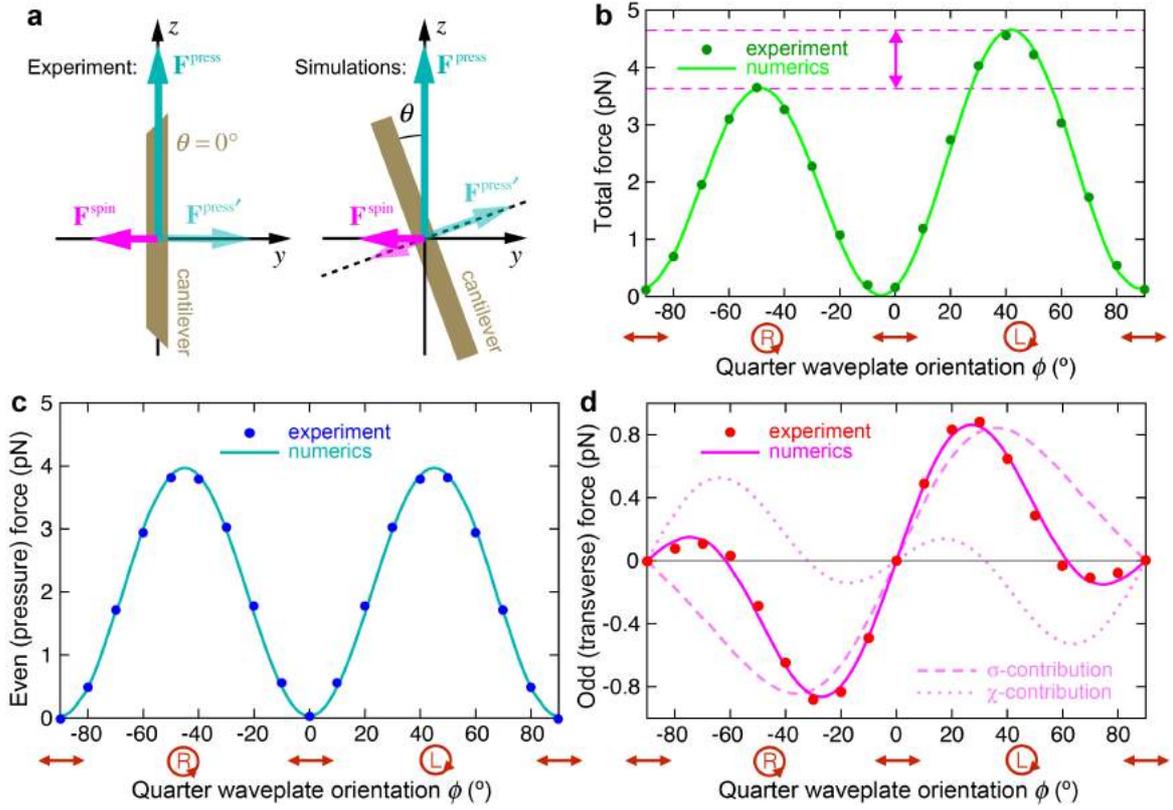

**Figure S3. (a)** Schematics of the mixing of the longitudinal radiation pressure with the transverse force in the field interaction with an $y \to -y$ asymmetric cantilever. Left: an asymmetric cantilever used in the experiment (see Fig. 2b). Right: symmetric but slightly rotated cantilever used in numerical simulations. **(b)** The total force $F(\phi)$ acting in the normal direction to the cantilever. Numerical simulations perfectly match the experimental measurement using the two fitting parameters: the cantilever thickness $d$ and its orientation angle $\theta$. **(c,d)** The $\phi$-even **(c)** and $\phi$-odd **(d)** parts of the total force $F(\phi)$ are associated with the longitudinal radiation pressure and weak transverse force, as shown in Fig. S2. The $\sigma$-dependent and $\chi$-dependent contributions to the transverse force, associated with the Belinfante spin momentum and imaginary Poynting vector are separately shown in **(d)**.

To verify that the $\tau$-dependent force in the experiment is indeed caused by the $y \to -y$ asymmetry of the cantilever, we note that it vanishes in numerical simulations with a perfectly-aligned symmetric cantilever: $\theta = 0$. To check the correspondence between the model asymmetry parameter $\theta$ and the $y \to -y$ asymmetry of the real cantilever, we also performed measurements with the same cantilever rotated by 180° about its vertical $x$-axis, which corresponds to the $y \to -y$ and $z \to -z$ transformations. Simultaneously, we changed the sign of $\theta$ in the numerical simulations, see Fig. S4a. The results, shown in Fig. S4, clearly demonstrate that the above transformations flip the even (radiation-pressure) part of the force, while leaving the odd (transverse) part of the force essentially unchanged. Some imperfections of these transformations between Fig. S3 and Fig. S4 are explained by the fact that these correspond to two independent experiments with re-assembling of the setup and also by the $z \to -z$ asymmetry of the real cantilever (Fig. 2b). These factors are taken into account by a new fitting procedure, which resulted in the value $\theta \simeq -0.1$ (i.e., $-5.6°$) for the "flipped" experiment in Fig. S4.

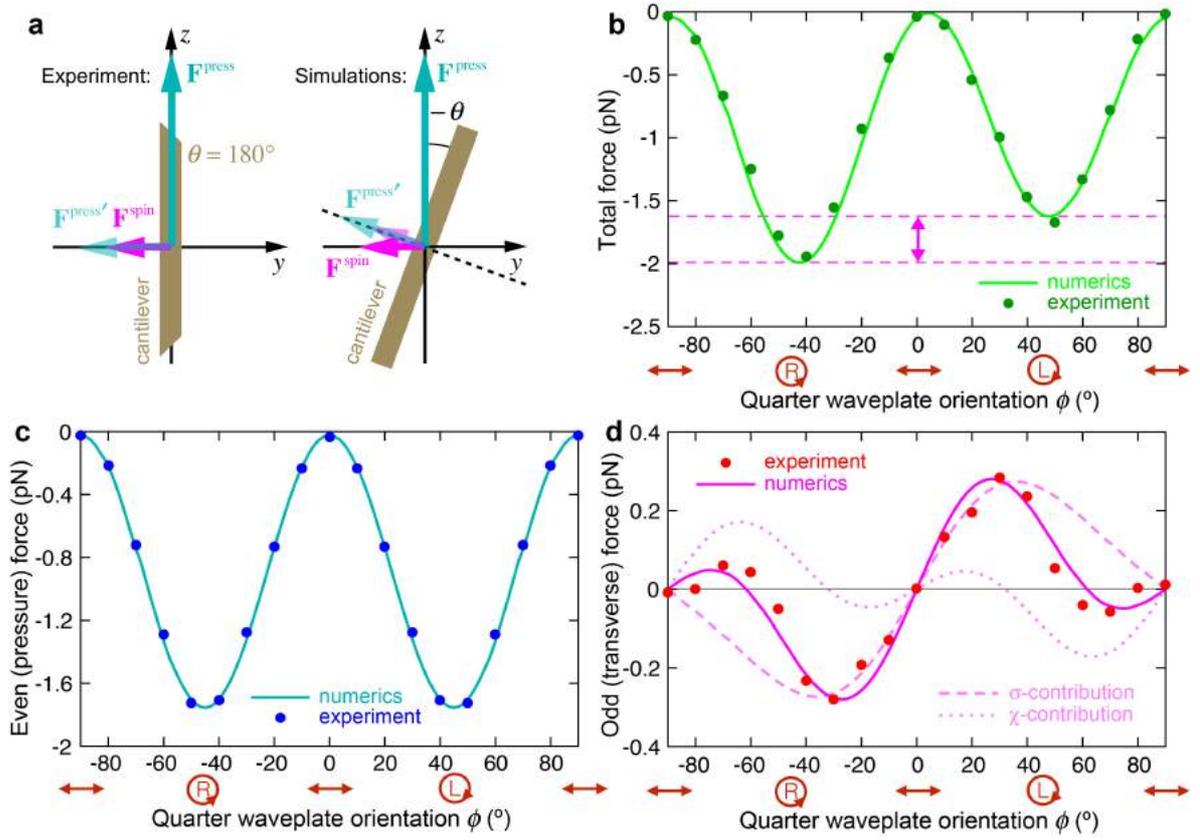

**Figure S4**. Same as in Fig. S3 but for the "flipped" cantilever: rotated by 180° about the $x$-axis in the experiment and oriented at the negative angle $-\theta$ in numerical simulations (cf. Fig. S3a and Fig. S4a). The flip of the $\phi$-even (radiation-pressure) part of the force proves that it is caused by the $y \to -y$ asymmetry of the cantilever. At the same time the $\phi$-odd (transverse) force remains almost unchanged, which proves that this is a robust field phenomenon.

Finally, using numerical simulations, we also investigated the dependences of the longitudinal and transverse forces on the *area* of the cantilever. This was done by calculating the forces at fixed QWP angle $\phi$ and varying the cantilever width $w$. To save simulation time, we modelled a thin cantilever with $d = 50$ nm, for which the imaginary-Poynting contribution to the transverse force is negligible. The results are shown in Fig. S5. One can see that the $\phi$-even force associated with the longitudinal radiation pressure grows near-*linearly* with $w$, which reflects its usual radiation-*pressure* nature related to the *planar surface* of the cantilever. In contrast to this, the $\phi$-odd force, associated with the transverse Belinfante spin momentum, approximately *saturates* after $w$ reaches few wavelengths. This means that the helicity-dependent force associated with the Belinfante spin momentum is *not a pressure force, but rather an edge effect related to wave diffraction on the vertical edges of the cantilever*. Indeed, one can show analytically that the transverse force *vanishes* for an infinite lamina without edges aligned with the $(x,z)$-plane: $F_y^{\text{spin}} = 0$. This is in extreme contrast to the *infinite* radiation-pressure force for the same lamina in the $(y,z)$-plane: $F_z^{\text{press}} = \infty$. In other words, this proves that the Belinfante spin momentum is indeed "virtual", and it does not exert the usual radiation pressure on planar objects. Nonetheless, it can be detected (as we do this in the present work) due to its weak interaction with the edges of finite-size probes.

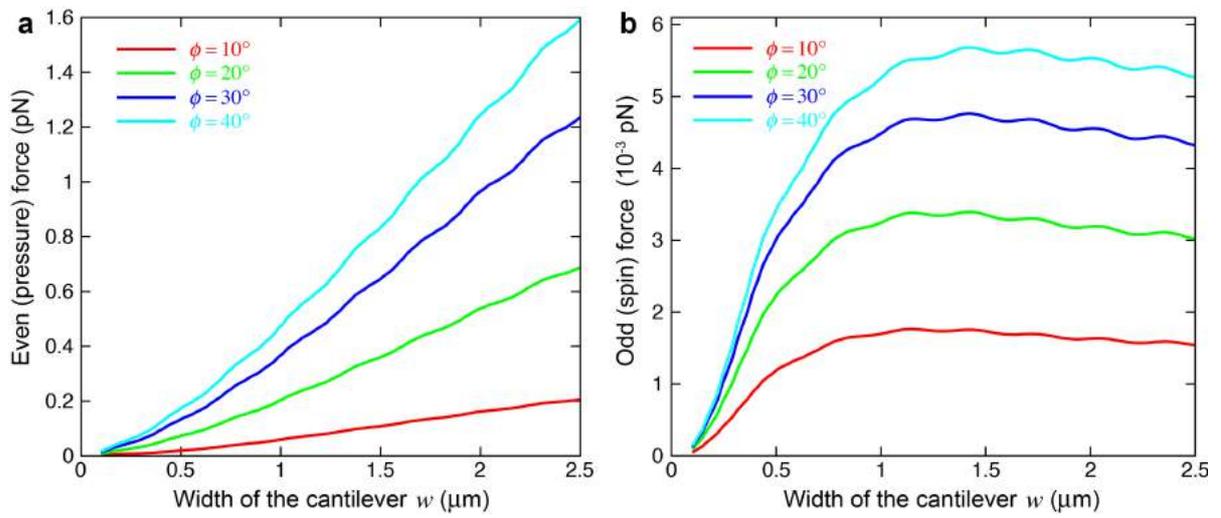

**Figure S5**. Numerically-calculated $\phi$-even **(a)** and $\phi$-odd **(b)** parts of the total force acting on the cantilever versus the cantilever width $w$. The linear growth of the even force reveals its radiation-pressure nature (proportional to the *area* of the cantilever surface). In contrast, the saturation of the odd (transverse) spin-dependent force indicates that it is produced by the field diffraction on the vertical *edges* of the cantilever.

## 3. Experimental measurements of optical forces using a nano-cantilever

Here we describe details of the experimental setup and measurements. In addition to the force measurements described in the main text and in Section 2 above, we performed a number of extra measurements of the field and cantilever properties in order to determine their parameters and control the system.

### *3.1. Characterization of the laser field*

The experimental setup is shown in Figure 2a. Two laser beams, orthogonal to each other, undergo a total internal reflection in a high numerical aperture (NA) objective lens and produce concentric evanescent areas above the glass cover slip of the objective (Fig. S6). The red laser 1 (Cube from Coherent Inc.) operates with a wavelength $\lambda = 660$ nm and power 50 mW. This radiation, with the polarization controlled using the QWP, is the source of the evanescent field under consideration. The green laser 2 (Versa-lase from Vortran Laser Technology Inc.) has a wavelength $\lambda' = 561$ nm and power 50 mW; its radiation is *p*-polarized at the glass-air interface where the evanescent field is formed. This radiation is used for measuring the cantilever position.

The evolution of the laser beams in the high-NA objective and generation of the evanescent wave is illustrated in Figure S6 (the laser-1 beam is shown). The input off-axial laser beam is focused in the back focal plane of the high-NA objective lens. The off-axis refraction at the objective lens forms a collimated beam propagating at the refraction angle $\alpha$, which depends on the displacement of the incident beam. At a certain off-axis distance, the condition for total internal reflection at the cover-slip surface is reached (for refractive index $n_1 = 1.5$, the critical angle of incidence is $\alpha_c = 41.8°$), and the evanescent field is generated in the air above the surface.

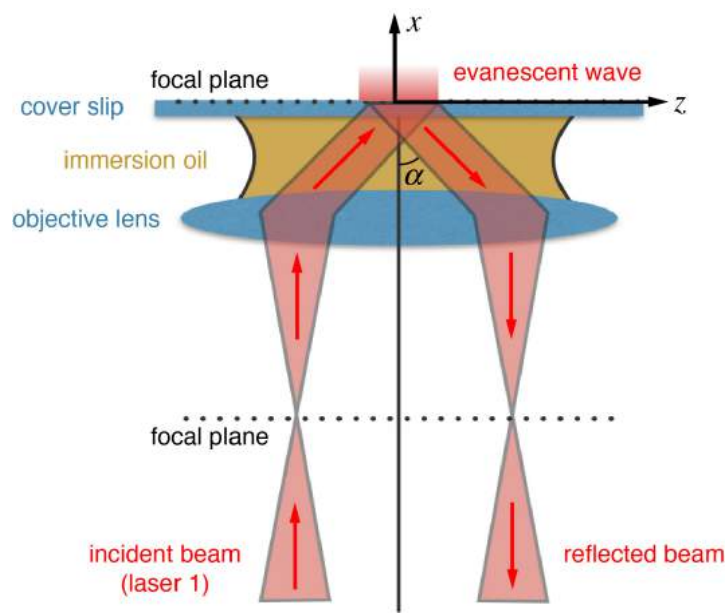

**Figure S6**. Formation of the evanescent field by the total internal reflection of the off-axis laser-1 beam in the high-NA objective covered by a glass slip. The laser-2 beam forms a similar concentric evanescent field in the orthogonal $(x,y)$-plane.

In this technique, we cannot measure the angle of incidence $\alpha$ directly, but it can be determined from the exponential decay length $\kappa^{-1}$ of the evanescent wave (i.e., the $x$-distance above the glass surface where the field amplitude drops by a factor of $e^{-1}$). We determined the decay length by measuring the displacement of the cantilever by the $p$-polarized laser 1 (which is proportional to the scattered intensity of the laser-1 radiation) as the cantilever tip is moved away from the surface, Fig. S7. Fitting the data with an exponential function results in a decay length $\kappa^{-1} = (150 \pm 4)$ nm. According to Eqs. (S14) and (S15), this corresponds to the angle of incidence $\alpha = 54.6° \pm 0.5°$.

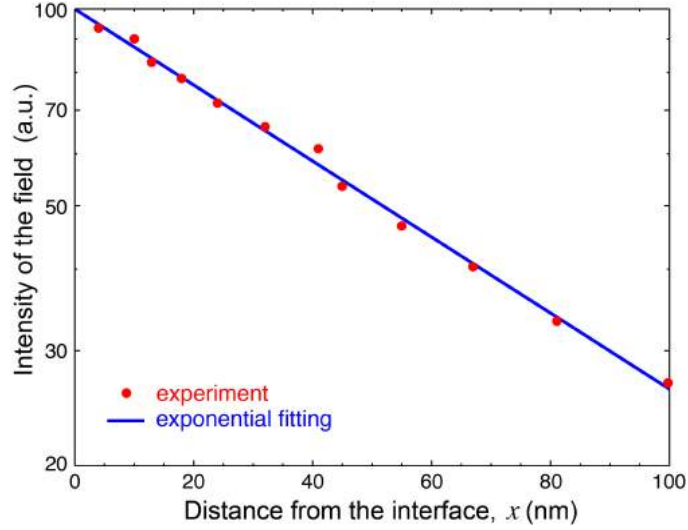

**Figure S7**. Measurement of the exponential decay of the evanescent-wave intensity (laser 1 here) above the glass surface.

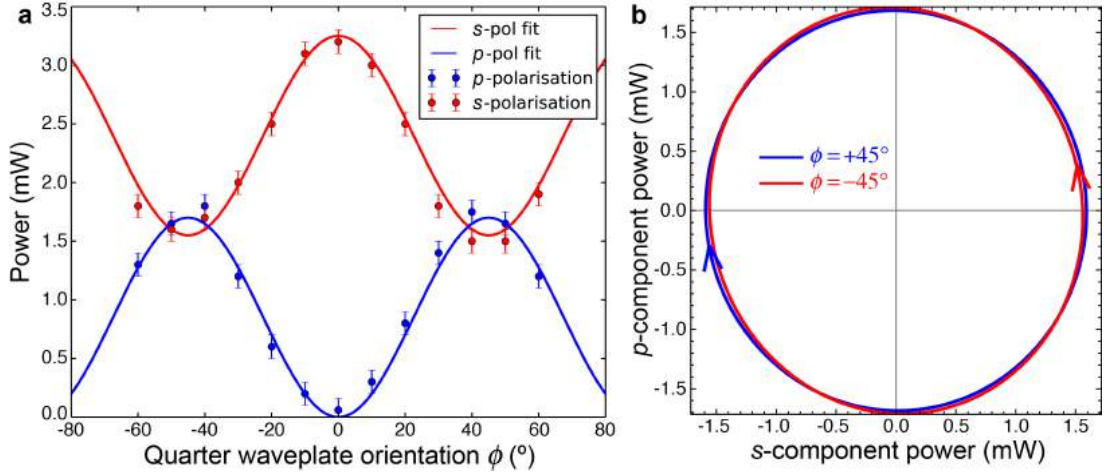

**Figure S8.** (a) The measured $s$ and $p$ components of the incident-beam polarization after the QWP and mirror. (b) Right-hand and left-hand circular polarizations are obtained for the QWP angles $\phi = \pm 45°$.

The polarization state of the laser-1 field is controlled by the rotating quarter waveplate (QWP) (Figs. 2a and S1). Its orientation is varied in the range of angles $-90° \leq \phi \leq 90°$. Here $\phi = 0°$ corresponds to the linear $s$-polarization of the beam. Figure S8 shows the measured polarizations of the beam after passing the QWP and mirror, just before entering the objective lens. In particular, for $\phi = \pm 45°$ the polarization becomes circular with good accuracy [ $m_1 = \mp i$ in Eq. (S13)]. The corresponding polarizations in the evanescent wave are slightly elliptical, which is taken into account in Eqs. (S16)–(S22).

### 3.2. Optical detection of the cantilever position

The cantilever, interacting with the evanescent laser-1 and laser-2 fields, is placed 30 nm above the cover-slip surface (Fig. 2a). It is perpendicular to the surface with an accuracy of 1º and is adjusted to be in the centre of the Gaussian-like area of the evanescent fields (diameter $\sim 50\,\mu m$) within $\pm 2\,\mu m$. The cantilever deflection is measured using the scattered evanescent wave (SEW) detection system [20] involving the laser-2 radiation. Note that the laser-2 field does not interfere with the laser-1 radiation and does not affect the cantilever interaction with the probed laser-1 field. Furthermore, the reflected laser-2 beam is the only signal reaching the quadrant photodetector, because two sets of filters are used to stop the reflected laser-1 beam (Fig. 2a). We have checked that the cross-talk between the two lasers at the detector is smaller than 0.1%.

We monitor the distance between the cantilever tip and the cover-slip surface using the total intensity of the scattered laser-2 light (the SUM signal produced by the photodetector). This distance is kept constant during the measurements by using a negative feedback loop. We adjust the bending direction of the cantilever (i.e., normal to its plane) to be perpendicular to the propagation $(x,z)$ plane of the laser-1 beam: $\theta = 0$ (Fig. 2d). However, the asymmetric shape of the cantilever (Fig. 2b) introduces the $y \to -y$ asymmetry to the system, which results in a nonzero radiation-pressure force in the transverse $y$ direction.

### 3.3. Characterization of the nano-cantilever

The optical forces investigated in this work were detected using different types of cantilevers developed in collaboration with NuNano ltd. The cantilevers were produced from extra low-stress $Si_3N_4$ film on a Si substrate. The thickness of the film determined the thickness of the cantilever and varied from 50 nm to 200 nm. The length of the cantilevers was adjusted to obtain similar stiffness $\sim 10^{-5}$ N/m. The width of the cantilever was kept constant, $w = 1000$ nm. The force magnitude is determined from the cantilever deflection once the stiffness of the cantilever is known. We determined the stiffness (spring constant) of the cantilever using the following two methods.

(i) We measured the thermal power spectral density (PSD) of the cantilever, $S(f)$ (where $f$ is the frequency), see Fig. S9 for an example of one such measurement. The measured PSD was fitted using the equation for the over-damped harmonic oscillator [S30]:

$$S(f) = \frac{k_B T}{b\pi^2 \left(f_c^2 + f^2\right)}, \quad (S30)$$

where $k_B$ is the Boltzmann constant, $T$ is the absolute temperature, $f_c = \gamma/(2\pi b)$ is the corner frequency of the PSD, and $b$ is the damping term. The resulting spring constant for the case shown in Fig. S9 is $\gamma = (2.1 \pm 0.2) \cdot 10^{-5}$ N/m.

(ii) By using the method based on the equipartition theorem [S31], the spring constant can be obtained once the mean square displacement $\langle x^2 \rangle$ of the thermally activated cantilever is measured. In this case, the spring constant is $\gamma = k_B T / \langle x^2 \rangle$. For the cantilever used in Fig. S9, this yields the spring constant $\gamma = (1.8 \pm 0.2) \cdot 10^{-5}$ N/m.

Thus, the two methods provide very similar results, so that we are confident that the spring constant can be determined within 15% of its actual value.

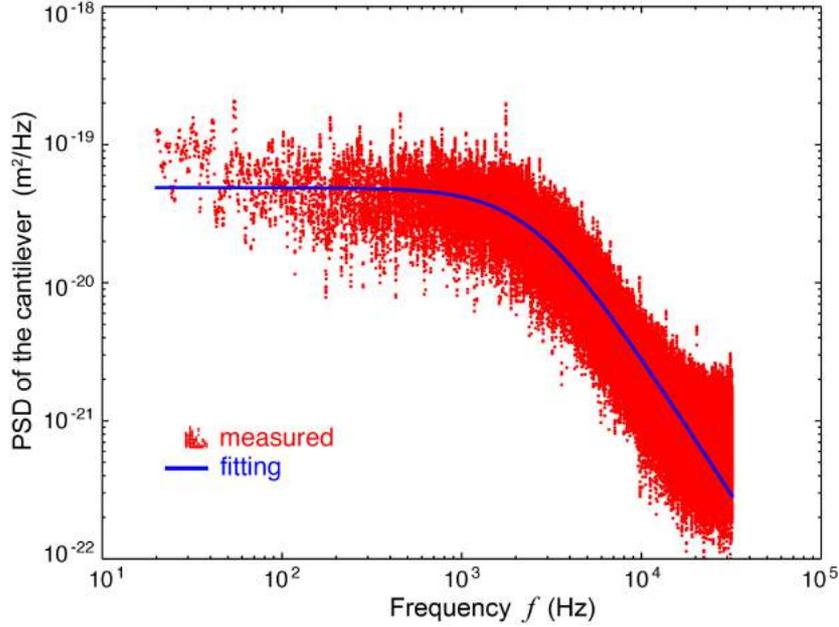

**Figure S9.** Experimentally-measured PSD of the thermally driven cantilever and its fitting curve obtained using the over-damped harmonic oscillator model (S30).

### *3.4. Optical force measurements*

Knowing the evanescent field properties, as well as the position and the stiffness of the cantilever, we can measure the desired optical forces from the cantilever deflections caused by the laser-1 field.

The intensity of the laser-1 beam is "on-off" modulated in time (TTL modulation) with a frequency of 1 Hz (see Fig. 3a). The deflection of the cantilever, $\Delta$, as well as the modulation signal are recorded at different QWP angles $\phi$, from −90º to +90º in steps of 10º. Thus, one complete set of measurements consists of 19 points, each measuring the position of the cantilever for 30 s. We did not notice any significant drift in the instrument during the 15 min necessary to complete a set of measurements.

The data are collected at 64 KHz but decimated by a factor of 100 to remove any cross correlation in the measurements. For each angle of the QWP orientation, the recorded trace is split into two sets of positions: one measured when the laser is "on" and the other one measured when the laser is "off". Two distributions of positions are generated with their relative mean and the standard error of the mean, as shown in Fig. 3a. The difference of the two mean values is the total displacement $\Delta(\phi)$ caused by the evanescent wave from laser 1. The deflection is finally multiplied by the spring constant to obtain the optical force normal to the cantilever plane: $F(\phi) = \gamma \Delta(\phi)$. To analyse the effects caused by the $y \to -y$ asymmetry of the cantilever, the measurements were repeated with the cantilever rotated by 180°, as shown in Figs. S3 and S4. To remove inessential force contributions, which do not vary with the QWP orientation (e.g., the gradient force emerging because the cantilever position is not exactly in the centre of the evanescent area, the possible mechanical action of the laser-2 radiation, etc.), the final data are offset to zero when the beam is *s*-polarized, i.e., when $\phi = 0$.

The results of the measurements of the optical force $F(\phi)$ are depicted in Fig. 3b. Most importantly, the dependence $F(\phi)$ is not symmetric with respect to $\phi \to -\phi$. This allows us to separate the $\phi$-even and $\phi$-odd components of the total force, which correspond to the longitudinal radiation-pressure effects and transverse forces. The measurement data for the retrieved forces are shown in Figs. 3c,d and S3, S4.

Thus, in our experiment and calculations, we have carefully traced and verified at all stages the appearance of the longitudinal and transverse optical forces. This allows us to unambiguously associate these measured forces with the longitudinal canonical momentum $P_z^{can}$, Eqs. (S6) and (S18), transverse Belinfante spin momentum $P_y^{spin}$, Eqs. (S7) and (S19), and the transverse imaginary Poynting momentum $\text{Im}\,\mathcal{P}_y$, Eq. (S20), in the evanescent wave [7].

### *3.5. Additional controls*

We note that the above measurements required an extremely low-noise environment without air currents around the probe. Therefore, air conditioning was switched off and a double enclosure was constructed around the LMFM unit.

We also checked that the thermally-induced deflection due to asymmetric illumination of the cantilever was negligible. The measurement protocol described above was repeated with the cantilever rotated by 180°, and it provided the same results within the experimental accuracy. These results confirm earlier findings [S32,S33] that the optical pressure on the cantilever is predominant over the photothermal effect.

Finally, we estimated the longitudinal torque $T_z$ on the cantilever produced by the spin angular momentum $S_z$ of the elliptically-polarized evanescent wave [7]. In principle, such torques could also cause a deflection proportional to the helicity of light. The ratio between the displacements due to the linear force $F_y^{spin}$ and the torque $T_z$ can be estimated as [S34]

$$\frac{(\Delta)_{F_y}}{(\Delta)_{T_z}} = \frac{2F_y^{spin}}{3T_z}l, \tag{S31}$$

where $l$ is the vertical length of the cantilever. A very conservative approximation, where $P_y^{spin}$ and $S_z$ are used in place of $F_y^{spin}$ and $T_z$, yields the ratio (S31) of ~350, confirming the negligible effect of the torque.

## Supplementary references